\documentclass[journal]{IEEEtran}
\ifCLASSINFOpdf
   \usepackage[pdftex]{graphicx}
\graphicspath{{Figures/}}
\DeclareGraphicsExtensions{.pdf,.jpeg,.jpg,.png}
\else
   \usepackage[dvips]{graphicx}
\graphicspath{{Figures/}}
   \DeclareGraphicsExtensions{.pdf,.jpeg,.jpg,.png}
\fi
\usepackage[outdir=./]{epstopdf}

\usepackage{amsmath}
\ifCLASSOPTIONcompsoc
  \usepackage[caption=false,font=normalsize,labelfont=sf,textfont=sf]{subfig}
\else
  \usepackage[caption=false,font=footnotesize]{subfig}
\fi
\DeclareMathOperator*{\argmax}{argmax}

\begin{document}
\title{Video-based Point Cloud Compression Artifact Removal}

\author{Anique~Akhtar,~\IEEEmembership{Student Member,~IEEE,}
		Wen~Gao,
        Li~Li,~\IEEEmembership{Member,~IEEE,}
        Zhu~Li,~\IEEEmembership{Senior~Member,~IEEE}
		Wei~Jia,~\IEEEmembership{Student Member,~IEEE,}
        Shan Liu,~\IEEEmembership{Senior~Member,~IEEE}% <-this % stops a space
\thanks{A. Akhtar, Z. Li, and W. Jia are with the Department of Computer Science and Electrical Engineering, 
University of Missouri-Kansas City, Kansas City, MO 64110 USA (e-mail: aniqueakhtar@mail.umkc.edu, zhu.li@ieee.org, wj3wr@umsystem.edu).}% <-this % stops a space
\thanks{W. Gao and S. Liu are with Tencent America, 661 Bryant St., Palo Alto, CA 94301 (e-mail: wengao@tencent.com, shanl@tencent.com).}
\thanks{L. Li is with the Department of Computer Science and Electrical Engineering, University of Missouri-Kansas City, Kansas City, MO 64110 USA, and also with the CAS Key Laboratory of Technology in Geo-Spatial Information Processing and Application System, University of Science and Technology of China, Hefei 230027, China (e-mail: lil1@ustc.edu.cn).}
}


\markboth{IEEE TRANSACTIONS ON MULTIMEDIA, 2021}%
{Akhtar \MakeLowercase{\textit{et al.}}: Video-Based Point Cloud Compression Artifact Removal}



\maketitle

\begin{abstract}
Photo-realistic point cloud capture and transmission are the fundamental enablers for immersive visual communication. The coding process of dynamic point clouds, especially video-based point cloud compression (V-PCC) developed by the MPEG standardization group, is now delivering state-of-the-art performance in compression efficiency. V-PCC is based on the projection of the point cloud patches to 2D planes and encoding the sequence as 2D texture and geometry patch sequences. However, the resulting quantization errors from coding can introduce compression artifacts, which can be very unpleasant for the quality of experience (QoE). In this work, we developed a novel out-of-the-loop point cloud geometry artifact removal solution that can significantly improve reconstruction quality without additional bandwidth cost. Our novel framework consists of a point cloud sampling scheme, an artifact removal network, and an aggregation scheme. The point cloud sampling scheme employs a cube-based neighborhood patch extraction to divide the point cloud into patches. The geometry artifact removal network then processes these patches to obtain artifact-removed patches. The artifact-removed patches are then merged together using an aggregation scheme to obtain the final artifact-removed point cloud. We employ 3D deep convolutional feature learning for geometry artifact removal that jointly recovers both the quantization direction and the quantization noise level by exploiting projection and quantization prior. The simulation results demonstrate that the proposed method is highly effective and can considerably improve the quality of the reconstructed point cloud.
\end{abstract}


% \fi
\IEEEpeerreviewmaketitle

\section{Introduction}
Recent significant advances in 3D sensors and capturing techniques have led to a surge in the usage of 3D point clouds in virtual reality/augmented reality (VR/AR) content creation and communications \cite{tulvan2016use}, as well as 3D sensing for robotics, smart cities, telepresence \cite{immersive}, and automated driving applications \cite{akhtar2019low}. A 3D point cloud can efficiently represent volumetric visual data such as 3D scenes and objects using a collection of discrete points with 3D geometry positions and other attributes (e.g., color, reflectance). Point cloud data offers advantages over polygonal meshes because it is more flexible and has real-time processing potential, since there is no need to process, store, or transfer surface topological information. With an increase in point cloud applications and improved capturing technologies, we now have high-resolution point clouds with millions of points per frame.

Based on their usage, point clouds can be categorized into point cloud scenes and point cloud objects. Point cloud scenes are dynamically acquired and are typically captured by LIDAR sensors. One example of a dynamic point cloud would be LIDAR sensors mounted atop a vehicle for mobile mapping and autonomous navigation purposes \cite{autonomous}. Point cloud objects can be further subdivided into static objects and dynamic objects. A static point cloud is a single object, whereas a dynamic point cloud is time-varying, where each instance of a dynamic point cloud is a static point cloud. Dynamic time-varying point clouds are used in AR/VR, volumetric video, and telepresence and can be generated using 3D models, i.e. CGI, or captured from real-world scenes using various methods such as multiple cameras with depth sensors surrounding the object and capturing movement over time. 

A volumetric video such as a dynamic point cloud provides an immersive media experience. A dynamic point cloud describes a 3D object using its geometry, respective attributes, as well as any temporal changes. Temporal information in the dynamic point cloud is included in the form of individual capture instances, much like 2D video frames. A dynamic point cloud can be viewed from any angle or viewpoint because it includes a complete 3D scene. This six degrees-of-freedom (6DoF) \cite{6DoF} viewing capability makes the dynamic point cloud essential for any AR or VR application. A single instance of a dynamic point cloud captured by 8i \cite{8i} could contain as many as one million points. Approximately 30 bits are used to represent the geometry (x,y,z), and 24 bits represent the color (r,g,b). The size of a single instance can be approximated as 6 Mbytes, which translates to a bitrate of 180 Mbytes per second without compression for a 30-fps dynamic point cloud. The high data rate is one of the main problems faced by dynamic point clouds, and efficient compression technologies to allow for the distribution of such content are still widely sought.

\begin{figure}[!t]
\centering
\subfloat[Higher bitrate]{\includegraphics[width=0.6\linewidth]{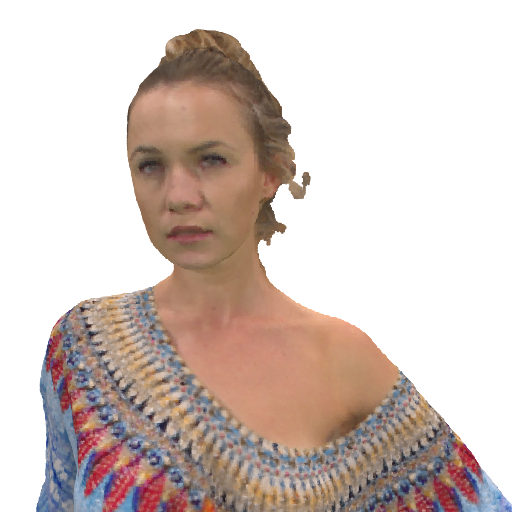}\label{highbits}}
\hfil
\subfloat[Lower bitrate]{\includegraphics[width=0.6\linewidth]{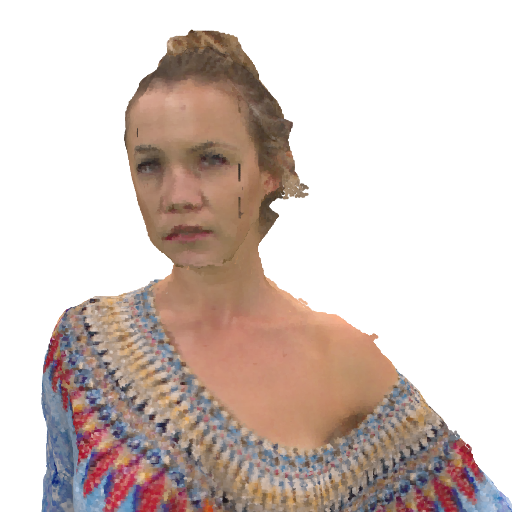}
\label{lowbits}}
\caption{Blocking effects in a point cloud coded at different bitrates using V-PCC encoding.}
\label{artifactblocking}
\end{figure}

The current state-of-the-art dynamic point cloud compression algorithm is the video-based point cloud compression (V-PCC) method \cite{VPCC2} which has been selected and developed for standardization by MPEG for dynamic point clouds. Under the V-PCC standard, a point cloud is first projected onto its bounding box patch by patch. Then, the patches are packed into a video for compression. During the video compression, the reconstructed geometry may suffer severe quality degradation due to the quantization errors. Blocking artifacts or compression artifacts are often introduced in compressed media due to distortion which is introduced by lossy compression techniques \cite{SOR}. The V-PCC coded point cloud yields excellent reproduction without noticeable artifacts at high or moderate bitrates. However, at low bitrates, the reconstructed point cloud suffers from visually annoying artifacts due to coarse quantization. Fig. \ref{artifactblocking} shows two versions of a point cloud encoded at different bitrates. As can be seen, there is no visible blocking artifact in the point cloud coded at a higher bitrate, while severe blocking artifacts exist in the one coded at a lower bitrate. Since blocking artifacts significantly degrade the visual quality of the reconstructed point cloud, it is desirable to identify these artifacts and remove them from the reconstructed point cloud.

This paper proposes the first deep-learning-based geometry artifact removal algorithm for the V-PCC standard for dynamic point clouds. Ours is a pioneering work in V-PCC artifact removal. The proposed framework offers the following contributions:
\begin{itemize}
\item We present a projection-aware 3D sparse convolutional neural network-based framework for point cloud artifact removal. Our sparse convolutional network learns an embedding and then regresses over this embedding to learn the quantization noise. Experimental results show that our method significantly improves the quality of the V-PCC reconstructed point cloud in terms of both objective evaluations and visual comparison.
\item We observe that the geometry distortion of the V-PCC reconstructed point cloud exists only in the direction of the V-PCC projection. We exploit this prior knowledge to learn both the direction and level of quantization noise by limiting the degree of freedom of the learned noise. We employ Chamfer distance as our loss function and use MSE-PSNR as our quality evaluation metrics.
\item We identify a \textit{patch correspondence mismatch problem} that arises due to a difference in the number of points in the original geometry and the V-PCC reconstructed geometry. To solve this, we propose a sampling and aggregation scheme using a cube-centered neighbor search algorithm to find a better correspondence between the reconstructed geometry (after V-PCC encoding) and the original geometry (before V-PCC encoding). The sampling and aggregation scheme makes our method scalable to larger point clouds since the framework is not dependent on the number of points in a point cloud.
\end{itemize}

\section{Background}
In 2017, MPEG issued a call for proposals on Point Cloud Compression (PCC) to target an international standard for PCC \cite{callPCC}. As a result of this call, multiple proposals were submitted to MPEG. Since then, MPEG has been evaluating and improving the performances of the proposed technologies. MPEG has selected two technologies for PCC: Geometry-based PCC (G-PCC) \cite{GPCC} for static point cloud data as well as for dynamically acquired LIDAR point cloud data, and video-based point cloud compression (V-PCC) \cite{VPCC} for dynamic content. G-PCC employs octree in its coding scheme, whereas V-PCC projects point clouds onto 2D surfaces and then uses state-of-the-art HEVC video encoding to encode dynamic point clouds. However, V-PCC does introduce compression artifacts, primarily when encoded with a low bitrate.

To the best of our knowledge, compression artifact removal in V-PCC has not been studied so far. However, compression artifact removal techniques and deblocking have been extensively studied in image and video coding. Since V-PCC also employs state-of-the-art HEVC video coding, there is potential to learn from the video compression artifact removal techniques and use them for V-PCC artifact removal. The current state-of-the-art compression artifact removal techniques are based on deep learning. There have been several previous studies using residual networks \cite{zhang2017beyond} and GANs \cite{galteri2017deep}, as well as efforts employing memory-based deep learning architecture \cite{tai2017memnet}. All of these works are limited to a single image and do not utilize information from previous frames. Recent works, however, have exploited temporal information in restoration tasks to improve video compression artifact removal \cite{lu2018deep, guan2019mfqe}.

Point cloud deep learning has been attracting increasing attention, especially in the last five years \cite{3dsurvey}. PointNet \cite{pointnet} was among the earliest deep learning architectures for point cloud learning and employed pointwise fully connected layers followed by max pooling. This architecture was further improved into PointNet++ \cite{pointnet++} by adding hierarchical learning that could learn local features with a better contextual scale. Wang et al. \cite{ocnn} proposed an octree-based CNN for 3D shape classification that performs 3D CNN operations on the octants of the octree data structure. PointCNN \cite{pointcnn} achieved state-of-the-art results using a convolutional neural network on raw 3D point clouds. To perform convolution on raw point clouds, PointCNN makes the input permutation invariant by learning a transformation matrix using fully connected layers. SparseConvNet \cite{sparseconvnet} by Facebook was among the first sparse convolutional neural networks that achieved state-of-the-art results on point clouds. SparseConvNet introduced submanifold sparse convolutions that exploited the sparse nature of point clouds and ensured that the convolutions would not ``dilate'' the data. Since sparse convolutions are memory-efficient and the data remain sparse throughout the network, deeper architectures can be used for point clouds. MinkowskiNet \cite{minkowski} is another such implementation that employs sparse convolutions for 3D point cloud learning. Recent works have also explored newer architectures for point cloud learning \cite{pointhop, valsesia2020learning}.

\begin{figure}[!t]
\hspace*{-3mm}  
\includegraphics[scale=0.68]{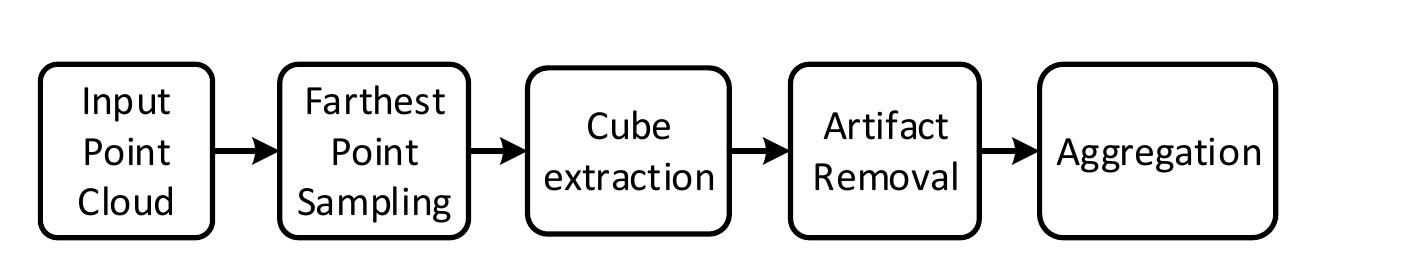}
\caption{System Model.} \label{system}
\end{figure}

The problem of point cloud denoising is an active research field which originated in the early 1990s. The works can be broadly divided into two categories: \textbf{optimization-based methods} and \textbf{deep-learning-based methods}. The optimization-based methods include techniques such as moving least squares (MLS)-based methods \cite{APSS, RIMLS}, locally optimal projection (LOP)-based methods \cite{LOP}, sparsity-based methods \cite{MRPCA,SOR}, non-local similarity-based methods \cite{BM3D}, and graph-based methods \cite{graph1,graph2}. However, the current state-of-the-art solutions are all deep learning-based methods. PU-Net \cite{yu2018pu} employed deep learning to learn multi-level features for point cloud denoising. This work was further improved to EC-Net \cite{yu2018ec}, which added edge-awareness to the network to further improve the results, especially around the edges. PU-GAN \cite{li2019pu} utilized generative adversarial networks (GANs) with patch-based learning for point cloud denoising. PUGeo-Net \cite{qian2020pugeo} incorporated discrete differential geometry into deep learning to learn underlying geometric structures from given sparse point clouds. These methods work well for synthetic noise (e.g. Gaussian noise), and some have even been tested on real-world noise introduced during point cloud capture. However, these methods are not optimized to work on compression artifact removal because of the nature of the quantization noise introduced during V-PCC.

Similarly, some work focuses on point cloud inpainting \cite{fu2018point,yu2020point,hu2019local}, wherein portions of point clouds lost during point cloud capture are completed. However, these methods do not work for compression artifact removal in V-PCC. In the last year, there has been notable work performed on deep learning solutions for point cloud compression \cite{deepcompression1, deepcompression2, deepcompression3, deepcompression4}. However, these solutions are still immature, and the standardized V-PCC is still widely used. There has also been some work conducted with respect to the improvement of the V-PCC standard \cite{li2020efficient}.

\begin{figure}[!t]
\centering
\subfloat[V-PCC projection onto "bounded box" planes. Image from \cite{VPCC1}]
{\includegraphics[width=0.47\linewidth]{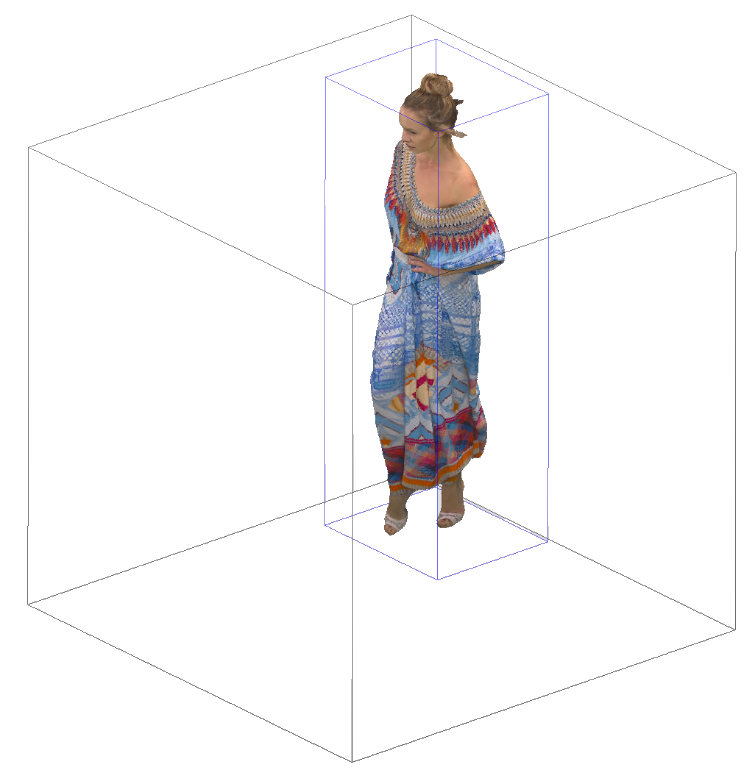}
\hfil
\includegraphics[width=0.5\linewidth]{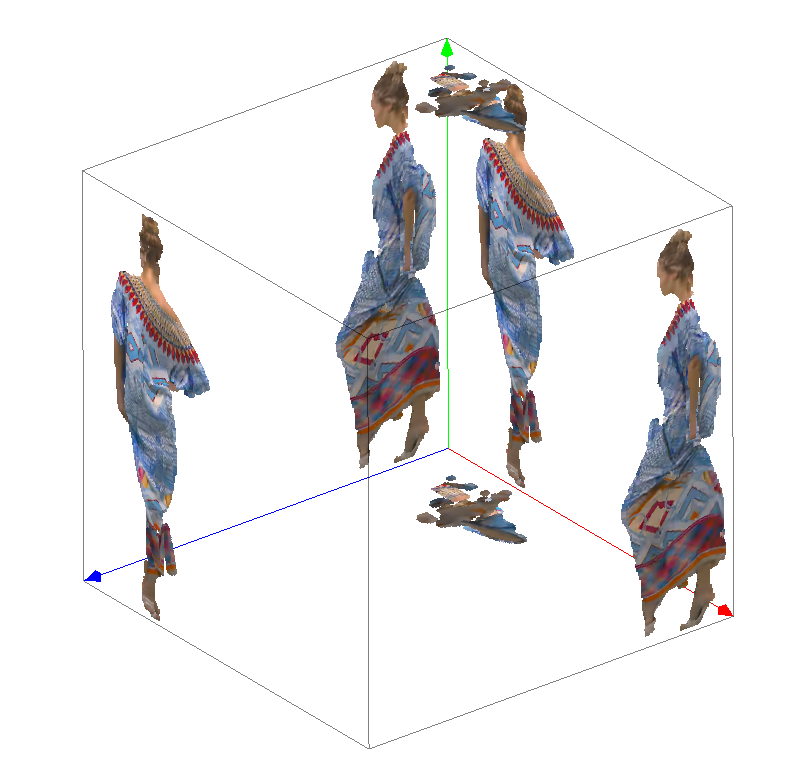}
\label{V-PCC-1}}
\vfil
\subfloat[V-PCC patch packing onto a 2D grid. Example of geometry (left)\\ and texture (right) images. Image from \cite{emerging}.]{\includegraphics[width=\linewidth]{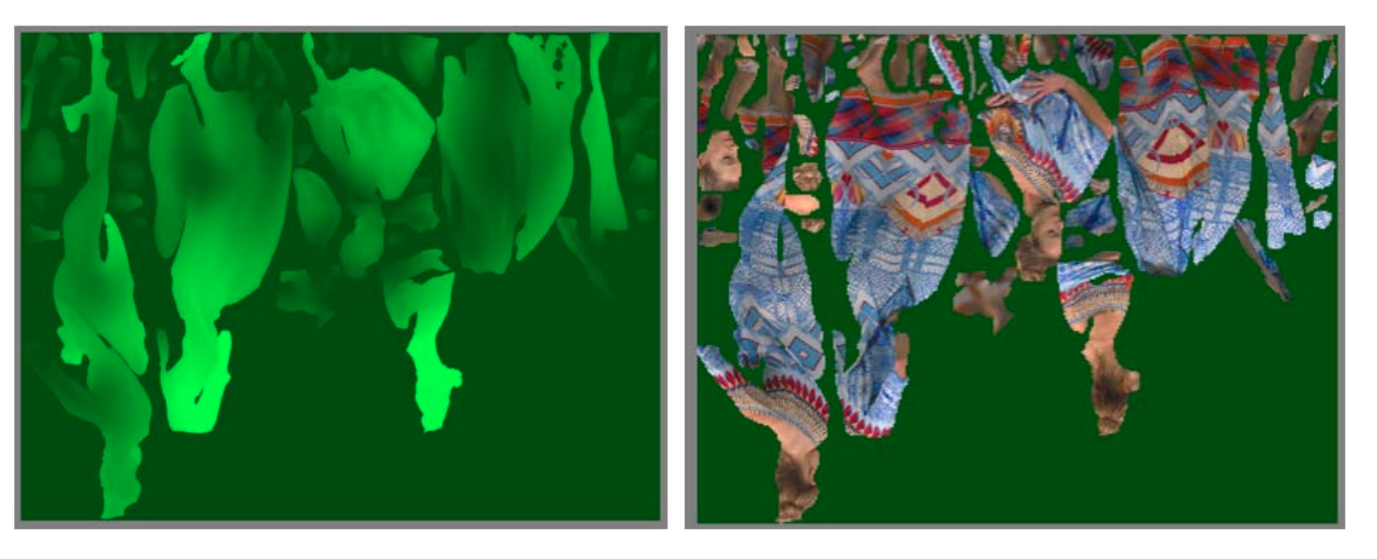}
\label{V-PCC-2}}
\caption{V-PCC: Example of 3D-to-2D projection. Although the figures show both geometry and texture information, in our work we are only concerned with geometry artifact removal.}
\label{V-PCC}
\end{figure}
 
\section{System Model}
As described earlier, the quantization noise from V-PCC coding can result in compression artifacts that can cause considerable degradation to the quality of the reconstructed point cloud. Our goal is to perform deep learning-based geometry artifact removal. We make our framework scalable to larger point clouds. We propose a sampling and aggregation scheme in which the point cloud is divided into smaller patches, and then these patches are passed through a geometry compression artifact removal network. Afterward, we combine the artifact-removed patches to form an artifact-removed point cloud. The system model is shown in Fig. \ref{system}. Taking advantage of the sparse nature of point clouds, we employ the sparse convolutional network \cite{sparseconvnet} which uses submanifold sparse convolutions \cite{graham2017submanifold} for our 3D learning.

\subsection{Problem Formulation}
V-PCC attempts to leverage existing video compression codecs for point cloud geometry and texture compression. V-PCC converts the point cloud into a set of different video sequences, one for geometry and one for texture information. In our work, we are only concerned with the noise introduced in the geometry compression. The video-generated bitstreams and the metadata needed to interpret these videos are then multiplexed together to generate the final point cloud V-PCC bitstream. V-PCC maps the input point cloud to a regular 2D grid by first decomposing the point cloud into a set of patches and then mapping these patches independently to a 2D grid using orthogonal projection. This process is shown in Fig. \ref{V-PCC-1}. V-PCC iteratively divides the point cloud into smaller patches to avoid auto-occlusions and to generate patches with smooth boundaries. To generate these patches, the normal for each point is first estimated. An initial clustering is obtained by associating each point to one of the six cube-oriented planes. More precisely, each point is associated with the plane that has the closest normal (i.e., the dot product of the point normal and the plane normal is maximum). This initial clustering is then refined by iteratively updating the cluster index by taking into account the point's normal and the neighboring point's cluster index. In this way, all the points in a refined patch are associated with a single plane. The majority of the points in the point cloud are projected to the cube plane that is closest to the normal of that point. This projection is only along one of the axes $(x, y,$ or $z)$. These patches are then projected onto the 2D grid using a process called packing to obtain a frame with texture and another with geometry. The final video sequence frames for geometry and texture are shown in Fig. \ref{V-PCC-2}.

\begin{figure}[!t]
\hspace*{-5mm} 
\centering
\subfloat[Gaussian Noise]{\includegraphics[width=0.51\linewidth]{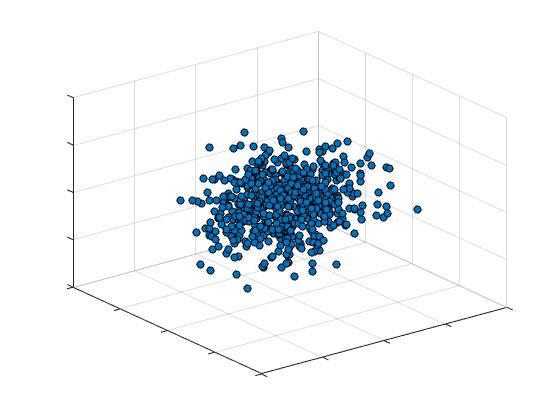}\label{fig_first_case}}
\hfil
\subfloat[V-PCC Quantization Noise]{\includegraphics[width=0.51\linewidth]{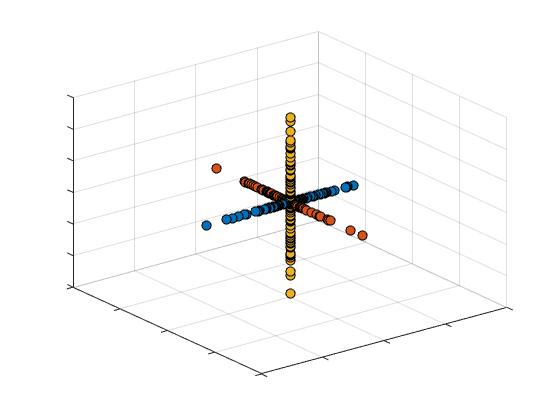}
\label{fig_second_case}}
\caption{Comparison of two different types of noises in a point cloud.}
\label{Noise}
\end{figure}

Afterward, these projected video frames are encoded by leveraging video compression techniques. Since these compression techniques are lossy, compression artifacts are often introduced due to quantization noise affecting the point cloud geometry. However, the artifact noise introduced in V-PCC geometry is only in one direction, as shown in Fig. \ref{Noise}. This is because each point is projected to only one plane, and therefore the artifact noise in that point is only in the direction of that plane. \textit{\textbf{This means that quantization noise introduced in each point is only along one of the axes (x, y, or z)}}. We leverage this property to learn the quantization noise level and quantization noise direction introduced by the V-PCC codec in the point cloud geometry. Since the quantization noise is along one of the axes, we make sure that our learned quantization noise for each point is also along a single axis. We exploit this prior knowledge of quantization noise direction to limit the degree of freedom of the learned quantization noise. We use the learned quantization noise to remove geometry artifacts from the reconstructed point cloud and improve its PSNR.

\begin{figure}[!t]
\centering
\subfloat[Dense surface]{\includegraphics[width=0.49\linewidth]{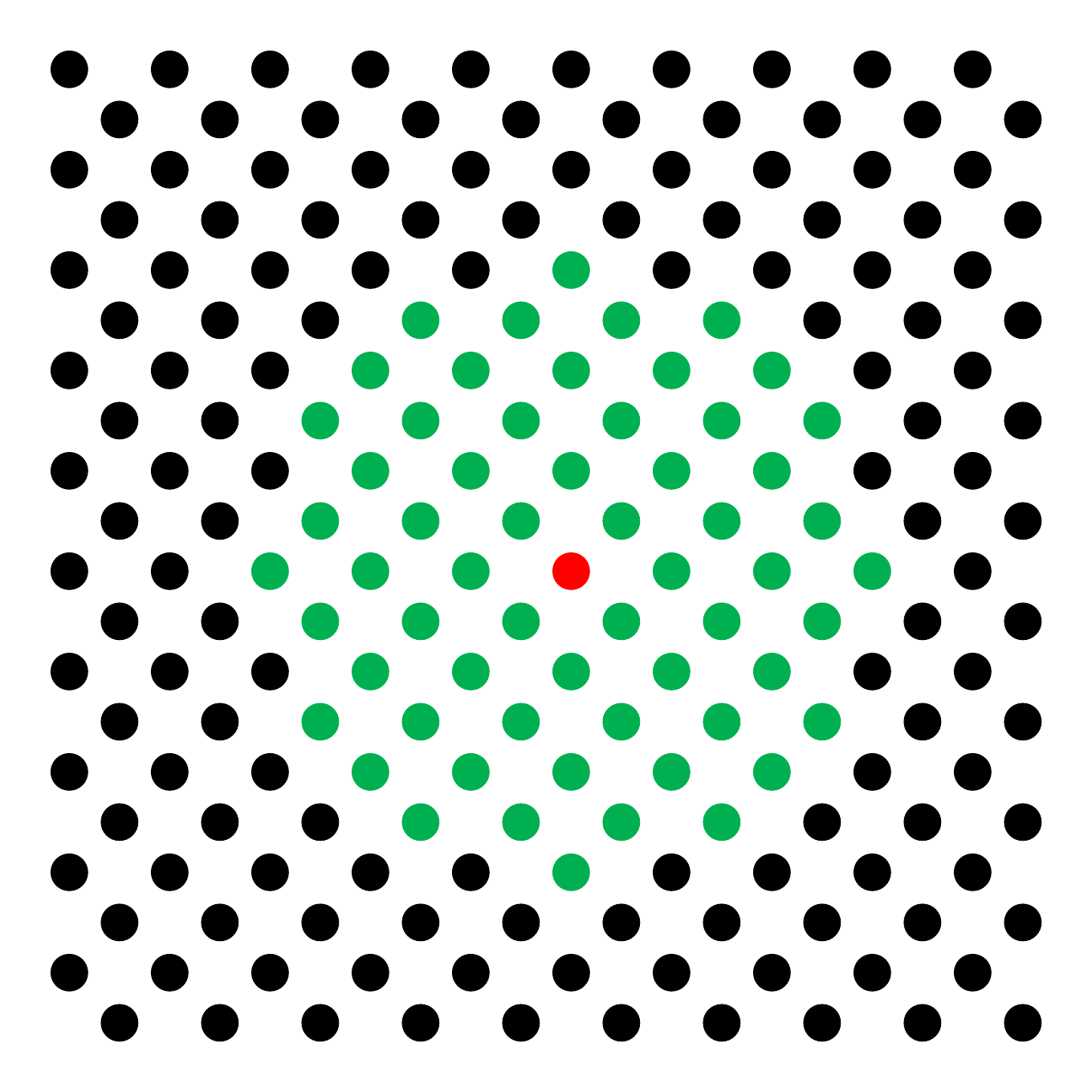}\label{points_1}}
\hfil
\subfloat[Sparse surface]{\includegraphics[width=0.49\linewidth]{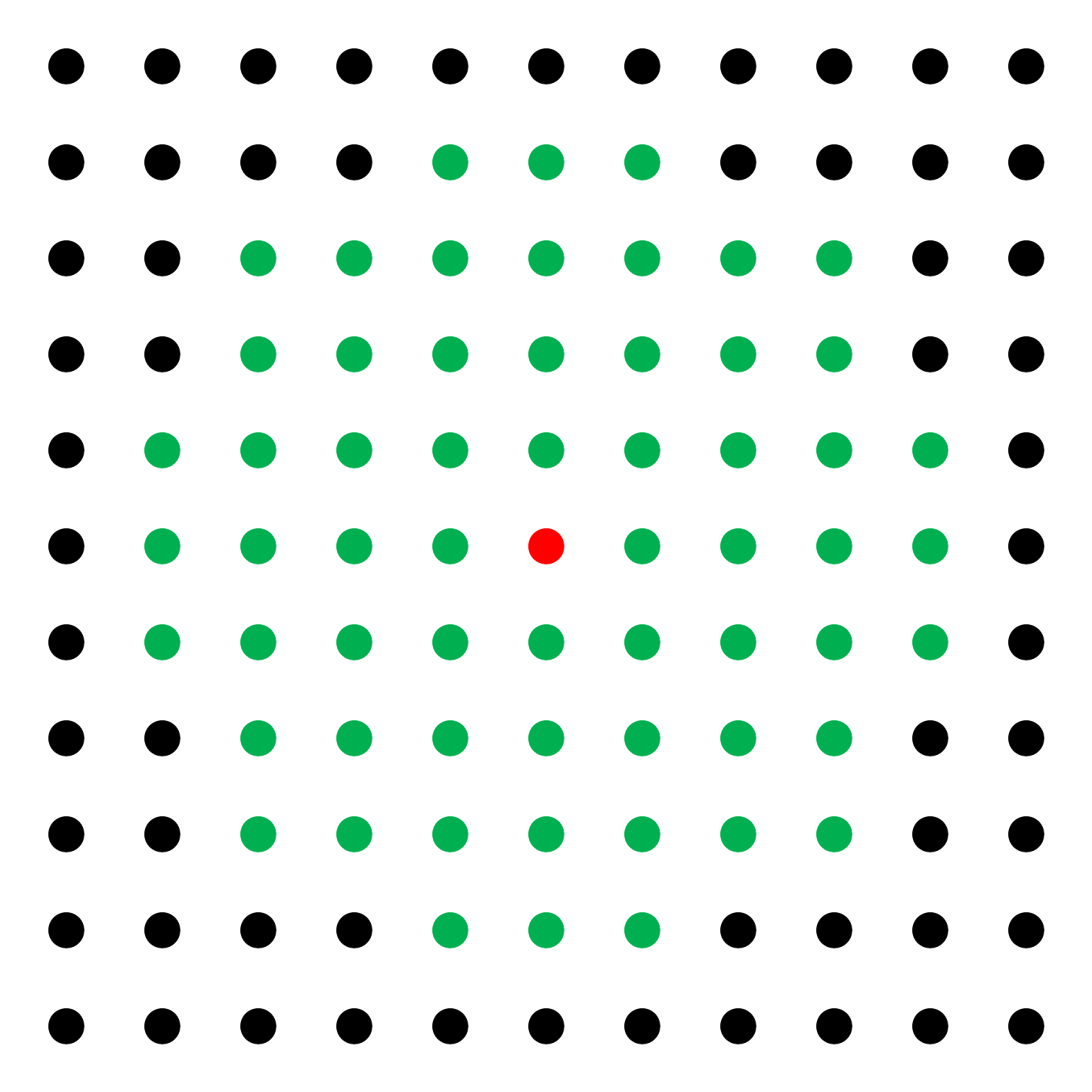}
\label{points_2}}
\caption{Patch correspondence mismatch problem for $k=61$. The k-NN search covers a smaller surface on a denser point cloud as compared to a sparser point cloud.}
\label{correspondence}
\end{figure}

\begin{figure}[!t]
\centering
\vspace*{-10mm}  
\includegraphics[scale=0.4]{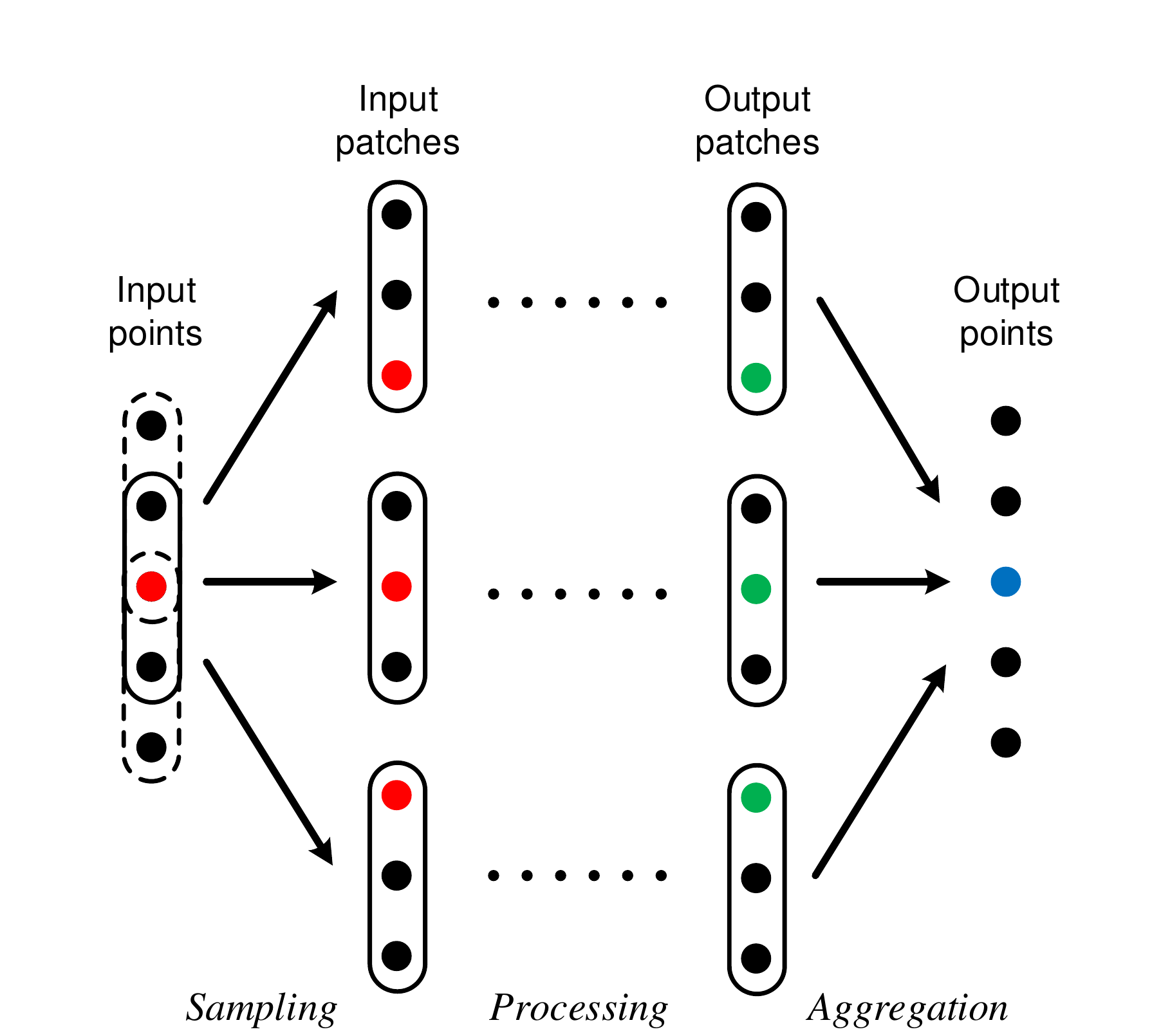}
\caption{An example sampling and aggregation scheme. The red point is the input point that is sampled into three patches. These three patches are processed to obtain three processed green points. The three green points are then aggregated to form the blue output point.} \label{aggregation}
\end{figure}

\begin{figure*}[!t]
\hspace*{-10mm}
\includegraphics[scale=0.65]{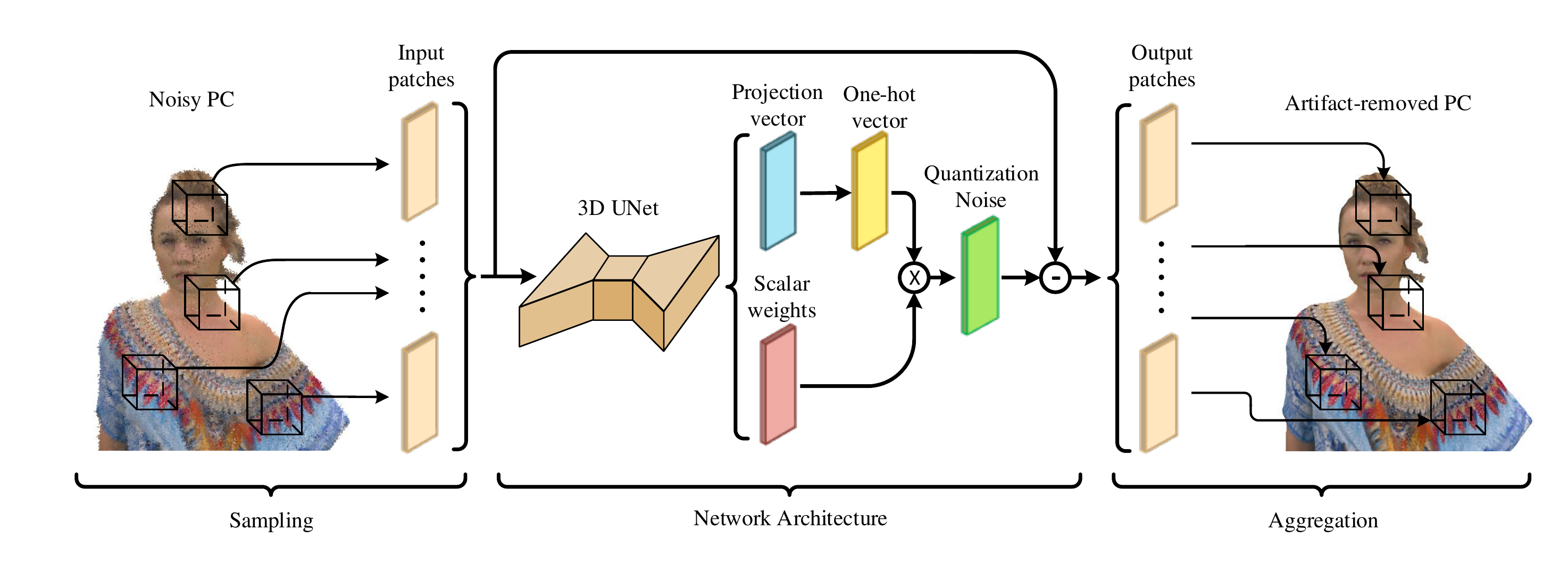}
\vspace*{-6mm}
\caption{Overview of the proposed point cloud artifact removal scheme. The input point cloud is divided into smaller patches that are fed into a sparse U-Net, which produces the projection vector and the scalar weights for each patch. The projection vector and the scalar weights are used to calculate the quantization noise that is then removed from the reconstructed point cloud patch. The output patches are then aggregated to obtain the artifact-removed point cloud.} \label{system2}
\end{figure*}

\subsection{Sampling}
The size of a point cloud can vary greatly, from point clouds with only a few thousand points to point clouds with millions of points. We propose a patch-based sampling and aggregation scheme to make our framework scalable to all sizes of point clouds. We sample a large point cloud into smaller neighborhood patches to ensure efficiency for practical application by offering affordable memory consumption on a cube basis, along with parallel processing. 

For each extracted patch, we need to find the patch in the original/ground-truth point cloud and the corresponding patch in the V-PCC reconstructed/noisy point cloud. A lossy low bitrate V-PCC reconstructed point cloud usually has fewer points than the original point cloud, consequently making the reconstructed point cloud sparser. Traditional patch-based point cloud deep learning models employ k-nearest neighbor (k-NN) search algorithms to obtain patches. However, when the number of points in the reconstructed point cloud differs from the number of points in the original point cloud, the k-NN search for extracting a neighborhood patch would not work because it would result in a different patch surface area for the two point clouds. Consider the example of $k=61$: using k-NN to extract a patch of 61 points would occupy a much larger area in a sparser point cloud as compared with a dense point cloud. We call this the \textit{patch correspondence mismatch problem} and show an example of it in Fig. \ref{correspondence}.

To solve the \textit{patch correspondence mismatch problem}, we propose a cube-centered neighborhood search algorithm wherein we extract all the points inside a fixed cube volume. We employ farthest point sampling (FPS) \cite{FPS} to sample points on the noisy point cloud and then extract cube patches around the sampled points. We obtain the noisy point cloud patch and the associated ground truth patch of the same cube volume extracted from the same location from both point clouds.

Farthest point sampling is employed to sample $N$ points over the point cloud, and a cube neighborhood around these points is used to extract neighborhood patches. $N$ patches are extracted using the formula:
\begin{equation}
	N =\frac{n * C}{k}
\end{equation}
where $n$ represents the total number of points in the point cloud, $k$ is the approximate average number of points in the neighborhood patch, and $C$ is a variable used to control the average number of overlapping patches per point. More points can be sampled by increasing the value of $C$, which would result in a larger average number of overlapping patches per point. Each of the sampled points is used as a center point for a cube and all the points inside the cube are extracted to form an input neighborhood patch. The geometry of the points inside the patch is zero-centered and then normalized between zero and the length of the cube side. These smaller input patches are fed to our 3D U-Net architecture as shown in Fig. \ref{system2}. Depending on the value of $C$, each point is sampled into multiple input patches and processed to obtain output patches. Therefore, we obtain multiple processed points for each point. We employ an aggregation scheme to merge the output patches to obtain the final output point cloud. An example of this can be found in Fig. \ref{aggregation}.

\subsection{Aggregation}
Once we have the artifact-removed output patches, we aggregate them back together to form the final point cloud. Post-processing is performed on the output patches for which the normalization is removed, and they are moved back to their original locations as shown in Fig. \ref{system2}. Depending on the value of $C$, we obtain many overlapping patches: therefore, each point receives multiple geometry values from different patches. Each input point is sampled into multiple patches, resulting in multiple clean point outputs for each input point. The geometries of these clean output points are mean aggregated to obtain the final clean output point. 

One example sampling and aggregation scheme is shown in Fig. \ref{aggregation}. The number of input points is $n=5$, the number of patches sampled is $N=3$, and each patch is of size $k=3$. For reference, we can examine the red input point. This input point is sampled into three patches and processed to obtain three processed green points. We then take the mean of the three green processed points to obtain the blue aggregated output point.

\begin{figure*}[!ht]
\hspace*{-6mm}
\centering
\includegraphics[scale=0.8]{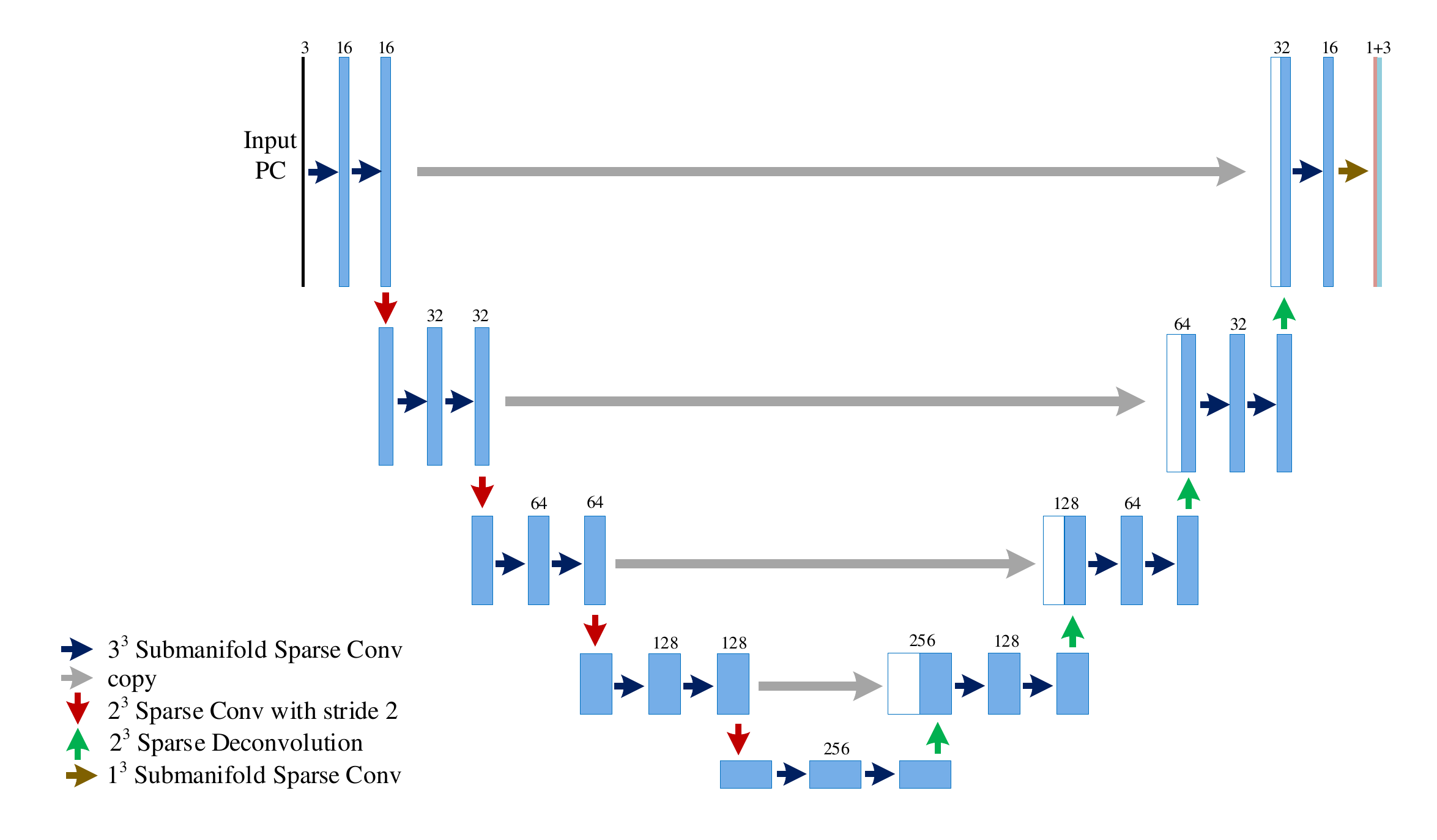}
\centering
\caption{3D U-Net Network Architecture.} \label{network}
\end{figure*}

\subsection{Network Architecture}
As explained in the previous section, we employ a cube-based patch sampling algorithm, so the number of points in the input patch varies. Traditional deep-learning-based point cloud processing networks work on a fixed number of input points to the network. We propose a fully convolutional 3D network that functions for variable input patch size. We employ sparse convolutional networks to create a fully convolutional network that offers the advantage of using a different number of points per patch, making the cube neighborhood patch extraction viable. Recall that our goal is to take in a 3D input patch of a V-PCC reconstructed point cloud and learn the per point quantization noise. The output of our 3D deep learning architecture is per point scalar weights and a per point projection vector. We use these to calculate the quantization noise.

\subsubsection{3D U-Net}
We employ submanifold sparse convolutions \cite{graham2017submanifold} that use a 3D sparse convolutional network \cite{sparseconvnet}. Sparse convolutions exploit the sparse nature of a point cloud and are much more memory efficient. Furthermore, sparse submanifold convolutions make sure the network does not ``dilate'' the sparse data and maintain the same sparsity level throughout the network. This helps us build and train deeper architectures like U-Net \cite{unet}.

U-Net architecture has been widely used in biomedical image segmentation tasks and usually employs 2D convolutions. We implemented a sparse convolution-based 3D U-Net architecture, the details of which are shown in Fig. \ref{network}. The architecture takes in a 3-dimensional geometry input patch, and the output is 4-dimensional: 3 dimensions for the projection vector and 1 dimension as a scalar weight. We use 3x3x3 $(3^3)$ submanifold sparse convolutions in each layer. We employ 2x2x2 $(2^3)$ convolutions with a stride of 2 for each \textit{downsampling}, whereas 2x2x2 $(2^3)$ deconvolution is used for \textit{upsampling}. The U-Net architecture typically consists of two paths: the encoder path and the decoder path. The encoder path captures the context of the point cloud, producing feature maps using strided convolutions to downsample. In the encoder path, with each downsampling, the number of points decreases but the feature dimension is doubled. The decoder path employs 3D deconvolution to upsample the point cloud. U-Net combines the information from the encoder path with that of the decoder path to obtain both the contextual information and localized information. U-Net only contains convolutional layers and does not employ any dense layers, making the network fully convolutional. This offers the added advantage of working with patches of variable input size.

\subsubsection{Quantization Noise Calculation}
The projection vector yields the direction of the quantization noise, whereas the scalar weight provides the level of the quantization noise. In V-PCC, a point is typically projected in one direction (x, y, or z); therefore, the quantization noise for each point is in a specific direction. Hence, it makes more sense to use a one-hot encoded projection vector for each point. We convert the projection vector into a one-hot vector by performing one-hot encoding using the maximum of the projection vector:

\begin{gather}
  Z_i(j) =
    \begin{cases}
      1 & \;\;\;\text{if $j = \argmax\limits_j V_i(j)$}\\
      0 & \;\;\;\text{else}
    \end{cases}
\end{gather}
Where $i$ is the point number, $j$ is the dimension, i.e. $j \in \{x,y,z\}$-axis, and $Z_i(j)$ is the one-hot vector, whereas $V_i(j)$ is the projection vector.

Once we have the one-hot vector, we multiply it with the scalar weight to learn the quantization noise. The per point quantization noise is then removed from the input patch to obtain an artifact-removed output patch. This is also illustrated in Fig. \ref{system2}. 

\subsubsection{Loss Function}
Our loss function is calculated by comparing the artifact-removed output patch to the ground truth patch. The loss function is applied to each patch before aggregation. We use Chamfer distance as the loss function in our architecture:
\begin{equation}
	L_{CD}(P_O,P_G) =\sum_{x\in P_O}\min_{y\in P_G} ||x-y||^2_2 + \sum_{y\in P_G}\min_{x\in P_O} ||x-y||^2_2
\end{equation}
Where $L_{CD}$ is the Chamfer distance loss function, $P_G$ is the ground truth patch and $P_O$ is the output artifact-removed patch calculated using input patch, projection vector, and scalar weights. Intuitively, the first term measures an approximate distance between each output point to the target surface, whereas the second term rewards even coverage of the output point cloud and penalizes any gaps.

%\begin{figure*}[!t]
%\centering
%\subfloat[Ground Truth]{\includegraphics[width=0.2\linewidth]{1_GT}}
%\hfil
%\subfloat[Bitrate br1]{\includegraphics[width=0.2\linewidth]{1_r6_input}}
%\hfil
%\subfloat[Artifact-removed PC: br1]{\includegraphics[width=0.2\linewidth]{1_r6_output}}
%\hfil
%\subfloat[Bitrate br2]{\includegraphics[width=0.2\linewidth]{1_r7_input}}
%\hfil
%\subfloat[Artifact-removed PC: br2]{\includegraphics[width=0.2\linewidth]{1_r7_output}}
%\hfil
%\subfloat[Bitrate br3]{\includegraphics[width=0.2\linewidth]{1_r8_input}}
%\hfil
%\subfloat[Artifact-removed PC: br3]{\includegraphics[width=0.2\linewidth]{1_r8_output}}
%\hfil
%\subfloat[Ground Truth]{\includegraphics[width=0.2\linewidth]{2_GT}}
%\hfil
%\subfloat[Bitrate br1]{\includegraphics[width=0.2\linewidth]{2_r6_input}}
%\hfil
%\subfloat[Artifact-removed PC: br1]{\includegraphics[width=0.2\linewidth]{2_r6_output}}
%\hfil
%\subfloat[Bitrate br2]{\includegraphics[width=0.2\linewidth]{2_r7_input}}
%\hfil
%\subfloat[Artifact-removed PC: br2]{\includegraphics[width=0.2\linewidth]{2_r7_output}}
%\hfil
%\subfloat[Bitrate br3]{\includegraphics[width=0.2\linewidth]{2_r8_input}}
%\hfil
%\subfloat[Artifact-removed PC: br3]{\includegraphics[width=0.2\linewidth]{2_r8_output}}
%\caption{Visual Results for all three bitrates.}
%\label{visual1}
%\end{figure*}

\begin{figure*}[!ht]
\hspace*{-2mm}
\centering
\includegraphics[scale=0.2]{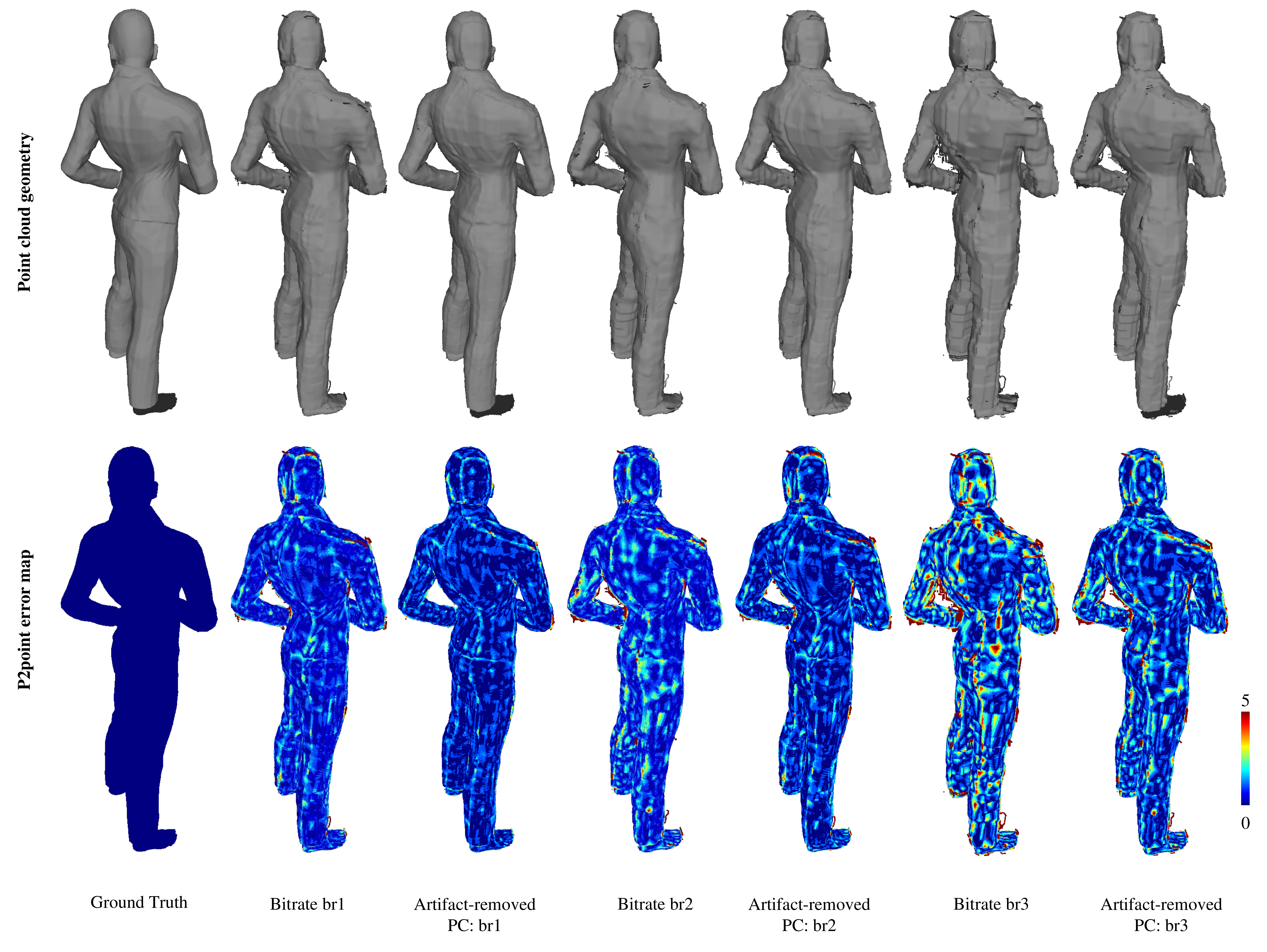}
\centering
\caption{Visual comparison of 'queen' showing ground truth, three different V-PCC reconstructed point clouds, and their corresponding artifact-removed point cloud. We show the geometry as well as point-to-point error map.} \label{Results_1}
\end{figure*}

\begin{figure*}[!ht]
\hspace*{-2mm}
\centering
\includegraphics[scale=0.2]{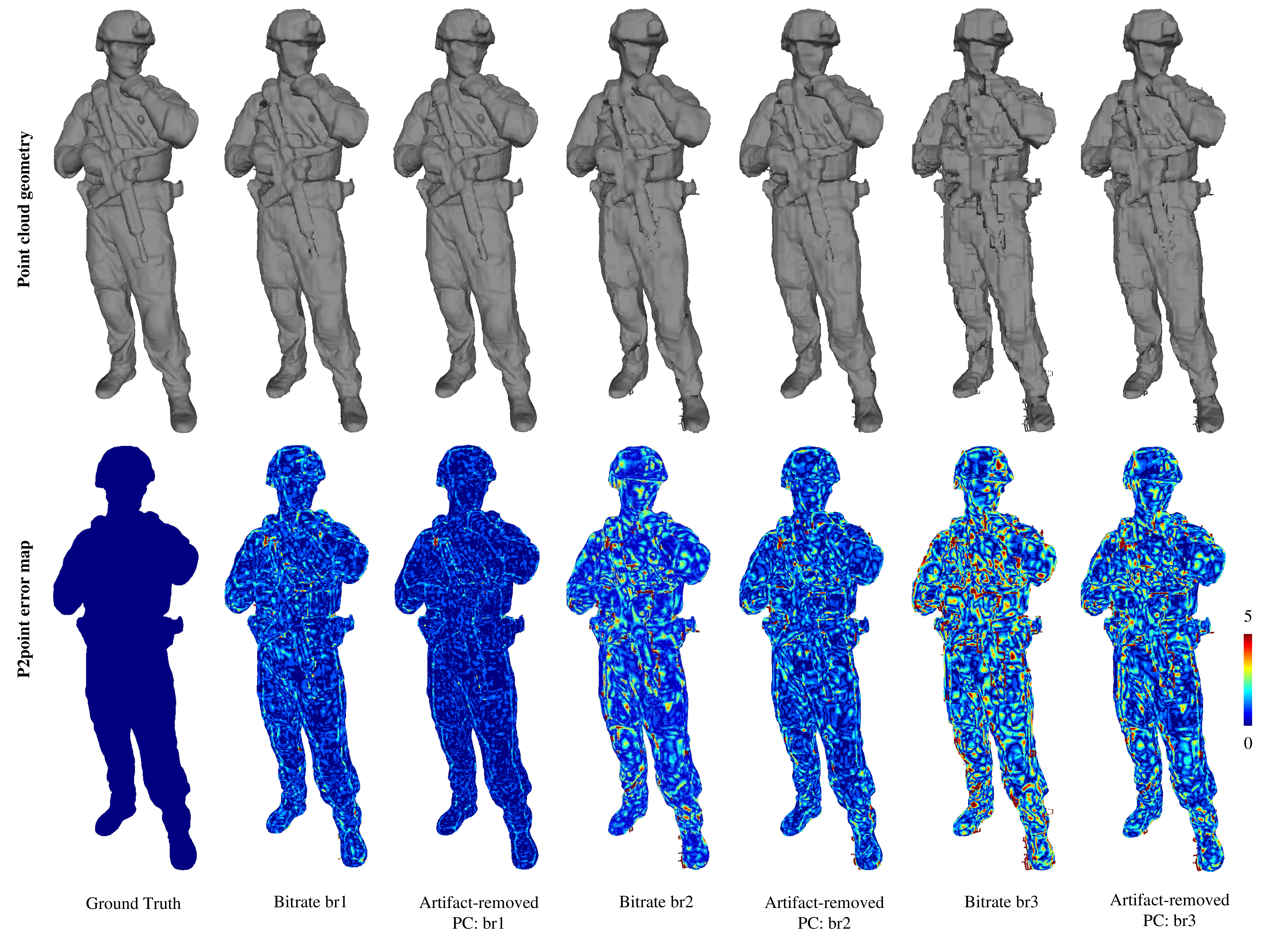}
\centering
\caption{Visual comparison of 'soldier' showing ground truth, three different V-PCC reconstructed point clouds, and their corresponding artifact-removed point clouds. We show the geometry as well as point-to-point error map.} \label{Results_3}
\end{figure*}

\section{Simulation Results}
We perform extensive simulations and show both the objective results and the visual comparison of our framework. Since this is the first work on V-PCC artifact removal, we have nothing with which to compare our work. However, we do show considerable improvement in the quality of the point cloud and perform multiple ablation studies to gain insight into the problem and explore alternative methods. For our simulation, we use the values of $k = 10000$ and $C = 20$.

\subsection{Dataset}
We use the 8i voxelized full bodies dataset from 8i labs \cite{8i}, which contains up to one million points per point cloud and is widely used by MPEG. The 8i dataset includes multiple sequences of point clouds. Each sequence has multiple point clouds representing 10 seconds of video captured at 30 fps for a total of 300 frames. We use two sequences for training (\textit{longdress, loot}) and three sequences for testing (\textit{redandblack, soldier, queen}). We use three different bitrates to encode these point clouds using V-PCC, shown in Table \ref{tab_bpp}. We label these bitrates as $br1$, $br2$, and $br3$ from the highest bitrate to the lowest bitrate, respectively.

\begin{table}[!htbp]
\renewcommand{\arraystretch}{1.3}
\caption{Different bitrates used in simulation}
\label{tab_bpp}
\centering
\begin{tabular}{c|c}
\hline
\textbf{Bitrate label} & \textbf{Actual bitrate}  \\
\hline
br1 & 0.01866 bpp \\
br2 & 0.01632 bpp \\
br3 & 0.01502\;bpp \\
\hline
\end{tabular}
\end{table}

\subsection{Objective Evaluation}
We use mean-squared-error (MSE) point-to-point PSNR (dB) as well as point-to-point Hausdorff PSNR (dB) as our objective metrics, which are calculated using MPEG's PC error tool \cite{PSNR}. We refer to MSE PSNR as simply PSNR in the rest of the section. We also measure the BD-Rate (Bj\"{o}ntegaard Delta Rate) \cite{bjontegaard2001calculation} improvement to determine how much savings our method achieves. The PSNR of the reconstructed point clouds obtained from the V-PCC encoder is measured before and after our artifact removal technique. The results are shown in Table \ref{tab_PSNR}. As can be determined for all three bitrates, our artifact removal technique considerably improves the PSNR as well as Hausdorff PSNR of the reconstructed point cloud.

We follow the MPEG common test condition to calculate the BD-rate using the PSNR metric. We compute the point-to-point distance for each frame of a sequence and obtain a total score by averaging across all frames. The final objective score was obtained by averaging across all three test sequences. We obtain a BD-rate savings of $11.3\%$. We also show the PSNR results for each 8i point cloud sequence separately for each bitrate in Table \ref{tab_PSNR_2}. We can also observe PSNR improvement for the bitrates for each point cloud sequence. From the results, we observe that higher PSNR improvement is achieved at a lower bitrate. This is because lower bitrates suffer from higher quantization noise, which our system can efficiently remove.

\begin{table}[!htbp]
\renewcommand{\arraystretch}{1.3}
\caption{Average PSNR results tested on three 8i sequences}
\label{tab_PSNR}
\centering
\begin{tabular}{c|cc|cc}
\hline
 & \multicolumn{2}{c|}{\textbf{PSNR (dB)}} & \multicolumn{2}{c}{\textbf{Hausdorff PSNR (dB)}} \\
\hline
\textbf{Bitrate} & \textbf{Noisy PC} & \textbf{Cleaned PC}  & \textbf{Noisy PC} & \textbf{Cleaned PC} \\
\hline
br3 & 59.62 & 60.47 & 37.02 & 37.21 \\
br2 & 61.84 & 62.36 & 39.58 & 39.64 \\
br1 & 64.20 & 64.53 & 41.15 & 41.20 \\
\hline
\hline
\multicolumn{3}{c}{BD-rate savings:} & \multicolumn{2}{l}{11.3 \%}\\
\hline
\end{tabular}
\end{table}

%\begin{table}[!htbp]
%\renewcommand{\arraystretch}{1.3}
%\caption{Simulation Results For Each Sequence}
%\label{tab_PSNR_2}
%\centering
%\begin{tabular}{c|c|cc}
%\hline
% & & \multicolumn{2}{c}{\textbf{PSNR (dB)}} \\
%\hline
%\textbf{Test PC} & \textbf{Bitrate} & \textbf{Noisy PC} & \textbf{Cleaned PC} \\
%\hline
%Queen & br3 & 60.23 & 60.79 \\
% & br2 & 62.35 & 62.90 \\
% & br1 & 64.95 & 65.21 \\
%\hline
% RenAndBlack & br3 & 59.44 & 60.38 \\
% & br2 & 61.62 & 62.18 \\
% & br1 & 63.90 & 64.27 \\
%\hline
% Soldier & br3 & 59.20 & 60.25 \\
% & br2 & 61.54 & 62.01 \\
% & br1 & 63.76 & 64.12 \\
%\hline
%\end{tabular}
%\end{table}

\begin{table}[!htbp]
\renewcommand{\arraystretch}{1.3}
\caption{Simulation Results For Each Individual Sequence}
\label{tab_PSNR_2}
\centering
\begin{tabular}{c|c|ccc}
\hline
 & & \multicolumn{3}{c}{\textbf{PSNR (dB)}} \\
\hline
\textbf{Test PC} & \textbf{Bitrate} & \textbf{Noisy PC} & \textbf{Cleaned PC}  &  \textbf{Improvement}\\
\hline
 Queen & br3 & 60.23 & 60.79 & 0.56\\
 & br2 & 62.35 & 62.90 & 0.55 \\
 & br1 & 64.94 & 65.21 & 0.27 \\
\hline
 RedAndBlack & br3 & 59.44 & 60.38 & 0.94\\
 & br2 & 61.62 & 62.18 & 0.56 \\
 & br1 & 63.90 & 64.27 & 0.37 \\
\hline
 Soldier & br3 & 59.20 & 60.25 & 1.05 \\
 & br2 & 61.53 & 62.01 & 0.48 \\
 & br1 & 63.76 & 64.12 & 0.36 \\
\hline
\end{tabular}
\end{table}

\subsection{Visual Results}
Visual results for point cloud artifact removal are shown in Fig. \ref{Results_1} and Fig. \ref{Results_3} for two different sequences: \textit{queen} and \textit{soldier}. We show the original (ground truth) point cloud, V-PCC reconstructed point cloud, and the artifact-removed point cloud for three different bitrates of V-PCC encoding. To visualize the point cloud, we first compute the normals for each point using 100 neighboring points; then, we set the shading to vertical and view the point cloud as a mesh. In this way, we can observe the point cloud geometry, which is more intuitive than vertex-color rendered images. However, since we use normals to visualize the point cloud, some points might appear black due to flipped normals, as shown from the right foot in Fig. \ref{Results_1}. We also plot the error map based on the point-to-point (P2point) D1 distance between decoded point clouds and ground truth to visualize the error distribution.

We can see that our method improves the quality of the point cloud, especially on the edges and surface of the point cloud. Although V-PCC performs well in quantitative objective comparison, its reconstructed point clouds contain noticeable artifacts when the bitrate is low. Our method removes these artifacts and considerably improves the visual quality of the point cloud. An interesting observation is that our artifact-removed point cloud compensates for some broken parts in the V-PCC reconstructed point cloud.

\subsection{Quantization Noise Calculation}
In this section, we compare our quantization noise calculation method with alternative methods. We use the V-PCC encoded bitrate of $br3$ for this experiment. Our current structure outputs 1-dimensional scalar weights and a 3-dimensional projection vector. We convert the projection vector to a one-hot encoded vector and multiply it by the scalar weights to calculate the quantization noise. After removing the quantization noise from the input patch, Chamfer loss is used to train the network. We label this \textbf{Our Method} and compare it with two alternative methods. \textbf{Method 1:} The network outputs 3-dimensional quantization noise that is directly removed from the input patch without any post-processing, and then Chamfer loss is used to train the network. \textbf{Method 2:} The network outputs 1-dimensional scalar weights and a 3-dimensional projection vector. Projection vectors are directly multiplied by the scalar weights to find quantization noise without converting them to a one-hot vector first. After removing the quantization noise from the input patch, Chamfer loss is used to train the network.

To summarize, in \textit{Method 1} the network directly outputs the quantization noise, whereas in \textit{Method 2} we remove the one-hot encoding part from our original architecture. The results of \textit{Our Method}, \textit{Method 1}, and \textit{Method 2} are compared in Table \ref{ablation_1}.

\begin{table}[!htbp]
\renewcommand{\arraystretch}{1.3}
\caption{Different quantization noise calculation methods}
\label{ablation_1}
\centering
\begin{tabular}{c|c|ccc}
\hline
 & & \multicolumn{3}{c}{\textbf{PSNR (dB)}} \\
\hline
\textbf{Test PC} & \textbf{Noisy PC} & \textbf{Our Method} & \textbf{Method 1} & \textbf{Method 2} \\
\hline
Queen  & 60.23 & \textbf{60.79} & 60.35 & 60.41 \\
RedAndBlack  & 59.44 & \textbf{60.38} & 59.92 & 60.02 \\
Soldier  & 59.20 & \textbf{60.25} & 59.76 &  59.93 \\
\textbf{Average}  & 59.62 & \textbf{60.47} & 60.01 & 60.12 \\
\hline
\end{tabular}
\end{table}

The results of Method 1 show that learning quantization noise directly from the network yields poor results. Similarly, comparison of our method with Method 2 shows that converting the projection vector to a one-hot vector before calculating the quantization noise considerably improves our results. This also shows that utilizing the prior knowledge of quantization noise which is only in the direction of the projection is beneficial to the learning of quantization noise.

\subsection{Sampling and Aggregation Schemes}
Currently, we employ the Farthest Point Sampling technique to sample points and extract a neighborhood around these points using cube extraction. Since there are overlapping neighborhood patches, we perform a mean aggregation scheme to obtain the final artifact-removed point cloud. We compare our method to a non-overlapping \textbf{octree-based} \cite{octree} cube division method. In the octree-based method, the point cloud is partitioned into cube nodes, and artifact removal is applied to each node. The differences between our method and the octree-based method are: 1) the patch sampling is performed using an octree; 2) there is no aggregation scheme, since the sampled patches are non-overlapping. The results of the comparison are shown in Table \ref{ablation_2}. We can observe from the results that our overlapping cubes-based sampling method substantially outperforms the octree-based sampling method.

\begin{table}[!htbp]
\renewcommand{\arraystretch}{1.3}
\caption{Different sampling and aggregation schemes}
\label{ablation_2}
\centering
\begin{tabular}{c|c|cc}
\hline
 & & \multicolumn{2}{c}{\textbf{PSNR (dB)}} \\
\hline
\textbf{Test PC} & \textbf{Noisy PC} & \textbf{Our Method} & \textbf{Octree-based} \\
\hline
Queen  & 60.23 & \textbf{60.79} & 60.38  \\
RedAndBlack  & 59.44 & \textbf{60.38} & 59.97 \\
Soldier  & 59.20 & \textbf{60.25} & 59.80 \\
\textbf{Average}  & 59.62 & \textbf{60.47} & 60.05 \\
\hline
\end{tabular}
\end{table}

\subsection{Choosing the value of C}
\begin{figure}[!t]
\hspace*{2mm}
\includegraphics[scale=0.235]{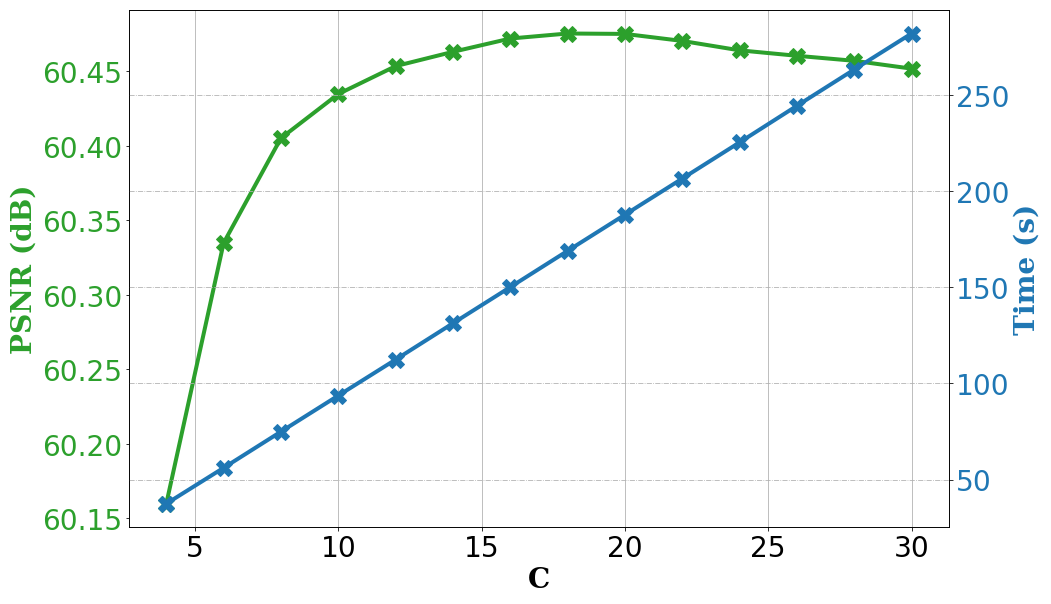}
\caption{PSNR (dB) and Time (s) complexity for different values of C.} \label{PSNR_C}
\end{figure}

As described earlier, $C$ is the variable used to control the average number of overlapping patches per point. A higher value of $C$ would result in a larger number of randomly sampled patches. To further study our sampling scheme, we perform a hyperparameter optimization experiment for the parameter $C$. We perform the simulation on the three test sequences of our 8i dataset (\textit{redandblack, soldier, queen}) and then plot the combined results. We vary the value of $C$ and measure the PSNR results as well as the computation time. Results of this experiment are shown in Fig. \ref{PSNR_C}. As can be observed, the PSNR increases with the value of $C$: PSNR is maximal at $C=18$ and starts to slightly decrease after that. The computation time is calculated as the average time to process a single 8i point cloud on an NVIDIA GeForce GTX 1080 Ti GPU. The computation time includes sampling, forward propagation through the network, and aggregation. As Fig. 11 shows, the computation time increases linearly with the value of $C$.

\section{Conclusion}
V-PCC encoding is the current state-of-the-art method for dynamic point cloud compression that has been selected by MPEG to be developed into a standard. However, quantization noise during V-PCC encoding results in severe quality degradation because it introduces compression artifacts. This paper presents a first-of-its-kind deep learning-based point cloud geometry compression artifact removal framework for V-PCC encoded dynamic point clouds. 
We leverage the prior knowledge that during V-PCC, the quantization noise is introduced only in the direction of the point cloud projection.
We employ a 3D sparse convolutional neural network to learn both the direction and the magnitude of geometry quantization noise. 
To make our work scalable, we propose a cube-centered neighborhood extraction scheme with a sampling and aggregation method to extract small neighborhood patches from the original point cloud. These patches are passed through the network to remove compression artifacts and then aggregated to form the final artifact-removed point cloud. Experimental results show that our method considerably improves the V-PCC reconstructed point cloud's geometry quality in both objective evaluations and visual comparisons.

\ifCLASSOPTIONcaptionsoff
  \newpage
\fi

\bibliographystyle{IEEEtrans}
\bibliography{Reference}



\vspace*{-2\baselineskip}
\begin{IEEEbiography}[{\includegraphics[width=1in,height=1.25in,clip,keepaspectratio]{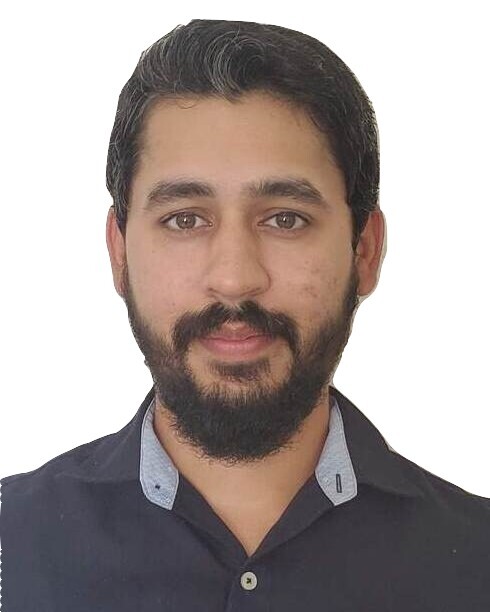}}]{Anique Akhtar} received his B.S. degree in Electrical Engineering from Lahore University of Management Sciences (LUMS), Lahore, Pakistan, in 2013, and the M.S. degree from Koc University, Istanbul, Turkey, in 2015. He is currently pursuing his Ph.D. at University of Missouri-Kansas City (UMKC). He has worked on wireless communication in the past and is currently working at the multimedia and communications lab at UMKC. His research interests include point cloud compression and communication, and deep learning solutions for point cloud processing.\end{IEEEbiography}

\vspace*{-1\baselineskip}

\begin{IEEEbiography}
[{\includegraphics[width=1in,height=1.25in,clip,keepaspectratio]{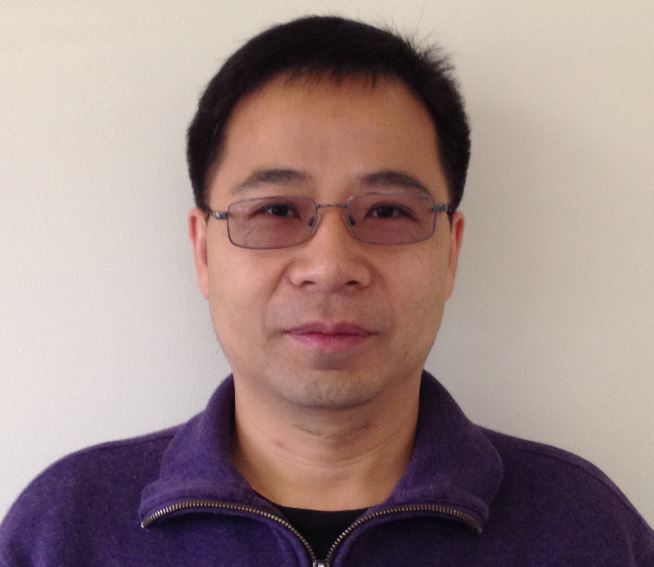}}]{Wen Gao} received the B.Eng. degree from University of Science and Technology of China, the M.S. degree from Peking University and Ph.D. degrees from Purdue University, all in Electrical Engineering. He is currently a Principal Researcher of Tencent Media Lab. He was formerly a senior staff software engineer/manager in Harmonic Inc, and a principal scientist/manager in Princeton Research Lab of Technicolor Inc. He has multiple years of research and development experience in video compression and wireless communications. He made contributions to variety of standard organizations such as MPEG, AVS, 3GPP, ATSC, DVB and IEEE802 standard committees, etc. His current research interests are in the areas of point cloud compression, video coding for machines and machine learning. 
\end{IEEEbiography}

\vspace*{-1\baselineskip}

\begin{IEEEbiography}
[{\includegraphics[width=1in,height=1.25in,clip,keepaspectratio]{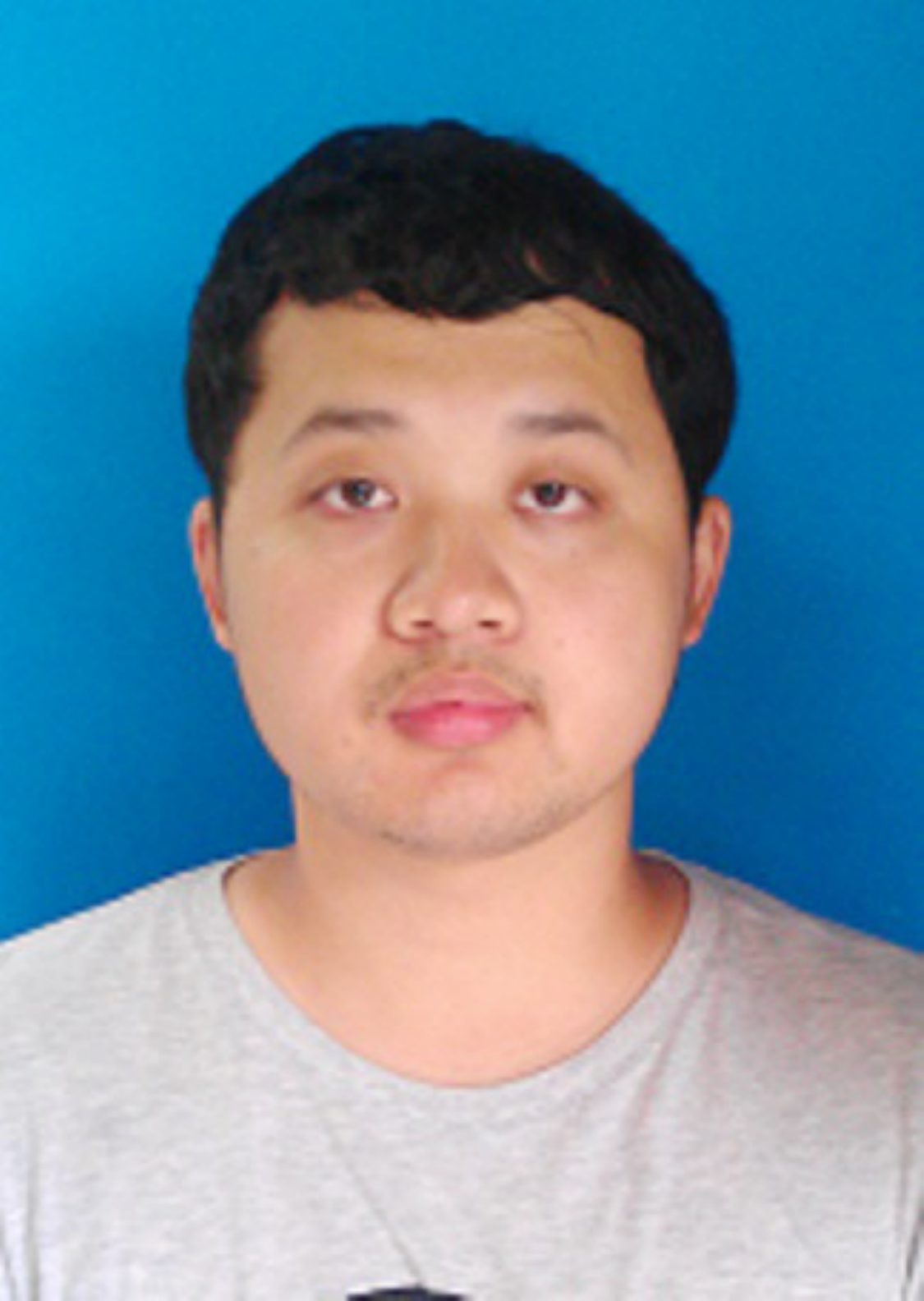}}]{Li Li} (M'17) is a research fellow with department of electronic engineering and information science, University of Science and Technology of China (USTC).
He received the B.S. and Ph.D. degrees in electronic engineering from USTC, Hefei, Anhui, China, in 2011 and 2016, respectively.
He was a visiting assistant professor in University of Missouri-Kansas City from 2016 to 2020.
His research interests include image/video coding and processing.
He received the Best 10\% Paper Award at the 2016 IEEE Visual Communications and Image Processing (VCIP) and the 2019 IEEE International Conference on Image Processing (ICIP).
\end{IEEEbiography}

\begin{IEEEbiography}
[{\includegraphics[width=1in,height=1.25in,clip,keepaspectratio]{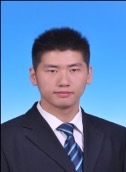}}]{Wei Jia} is a Ph.D. candidate with department of computer science electrical engineering, University of Missouri-Kansas City (UMKC).
He received the B.S. and M.S. degrees in electronic engineering from Beijing University of Posts and Telecommunications (BUPT), Beijing, China, in 2009 and 2012, respectively.
His research interests include image/video coding and processing, video-based point cloud compression, and deep learning.
\end{IEEEbiography}

\vspace*{-1\baselineskip}

\begin{IEEEbiography}
[{\includegraphics[width=1in,height=1.25in,clip,keepaspectratio]{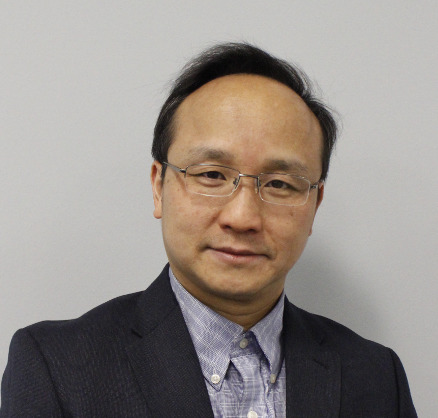}}]{Zhu Li} is now an Associate Professor with the Dept of Computer Science \& Electrical Engnieering (CSEE), University of Missouri,Kansas City, and director of the NSF I/UCRC Center for Big Learning (CBL) at UMKC. He received his PhD in Electrical \& Computer Engineering from Northwestern University, Evanston in 2004. He was AFOSR SFFP summer visiting faculty at the US Air Force Academy (USAFA), 2016 , 2017 , 2018 and 2020. He was Sr. Staff Researcher/Sr. Manager with Samsung Research America's Multimedia Standards Research Lab in Richardson, TX, 2012-2015, Sr. Staff Researcher with FutureWei (Huawei) Technology's Media Lab in Bridgewater, NJ, 2010~2012, an Assistant Professor with the Dept of Computing, The Hong Kong Polytechnic University from 2008 to 2010, and a Principal Staff Research Engineer with the Multimedia Research Lab (MRL), Motorola Labs, from 2000 to 2008. His research interests include point cloud and light field compression, graph signal processing and deep learning in the next gen visual compression, image processing and understanding. He has 47 issued or pending patents, 160+ publications in book chapters, journals, and conferences in these areas. He is an IEEE senior member, Associate Editor-in-Chief for IEEE Trans on Circuits \& System for Video Tech, associated editor for IEEE Trans on Image Processing(2020~), IEEE Trans.on Multimedia (2015-18), IEEE Trans on Circuits \& System for Video Technology(2016-19). He serves on the steering committee member of IEEE ICME (2015-18), he is an elected member of the IEEE Multimedia Signal Processing (MMSP), IEEE Image, Video, and Multidimensional Signal Processing (IVMSP), and IEEE Visual Signal Processing \& Communication (VSPC) Tech Committees. He is program co-chair for IEEE Intâ€™l Conf on Multimedia \& Expo (ICME) 2019, and co-chaired the IEEE Visual Communication \& Image Processing (VCIP) 2017.  He received the Best Paper Award at IEEE Int'l Conf on Multimedia \& Expo (ICME), Toronto, 2006, the Best Paper Award (DoCoMo Labs Innovative Paper) at IEEE Int'l Conf on Image Processing (ICIP), San Antonio, 2007.
\end{IEEEbiography}

\newpage
 
\begin{IEEEbiography}
[{\includegraphics[width=1in,height=1.25in,clip,keepaspectratio]{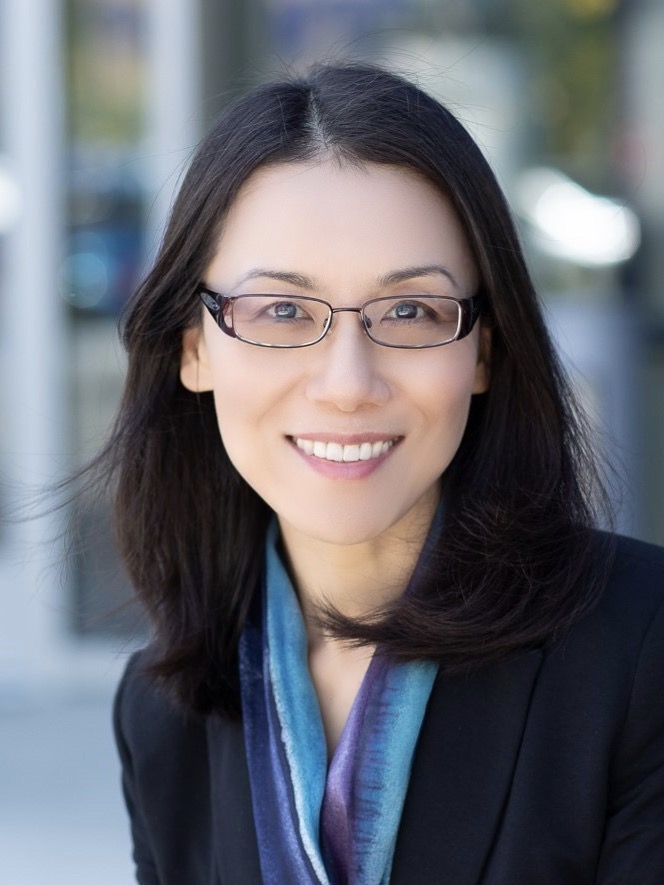}}]{Shan Liu} received the B.Eng. degree in Electronic Engineering from Tsinghua University, the M.S. and Ph.D. degrees in Electrical Engineering from the University of Southern California, respectively. She is currently a Tencent Distinguished Scientist and General Manager of Tencent Media Lab. She was formerly Director of Media Technology Division at MediaTek USA. She was also formerly with MERL and Sony, etc. Dr. Liu has been actively contributing to international standards since the last decade and served co-Editor of H.265/HEVC SCC and H.266/VVC. She has numerous technical proposals adopted into various standards, such as VVC, HEVC, OMAF, DASH, MMT and PCC, etc. Technologies and products developed by her and under her leadership have served hundreds of millions of users. Dr. Liu holds more than 200 granted US and global patents and has published more than 100 journal and conference papers. She was in the committee of Industrial Relationship of IEEE Signal Processing Society (2014-2015). She served the VP of Industrial Relations and Development of Asia-Pacific Signal and Information Processing Association (2016-2017) and was named APSIPA Industrial Distinguished Leader in 2018. She is on the Editorial Board of IEEE Transactions on Circuits and Systems for Video Technology (2018-present) and received the Best AE Award in 2019 and 2020, respectively. She also served or serves as a guest editor for a few T-CSVT special issues and special sections. She has been serving Vice Chair of IEEE Data Compression Standards Committee since 2019. Her research interests include audio-visual, volumetric, immersive and emerging media compression, intelligence, transport and systems.
\end{IEEEbiography}

\end{document}